\documentclass{article}
\usepackage{amssymb}
\usepackage{amsmath}

\input{tcilatex}

\begin{document}

\title{On the black holes in external electromagnetic fields.}
\author{V. A. Belinski \\
International Network of Centers for Relativistic Astrophysics \ \ \\
(ICRANet), 65122 Pescara, Italy.\\
Institut des Hautes Etudes Scientifiques (IHES), F-91440 \ \\
Bures-sur-Yvette, France.\\
\\
\ \ }
\date{}
\maketitle

\begin{abstract}
This preprint mainly reflects the new chapter we are prepearing for the
second edition of our book "Gravitational Solitons" (V. Belinski and E.
Verdaguer). However, here it is written in the form of more or less
self-consistent paper dedicated to the mathematical theory of black holes
immersed in the external electromagnetic field. The purpose of this
development is to describe the procedure by which one can construct a number
of axially symmetric black holes or naked singularities arranged along the
symmetry axis and immersed in an (also axially symmetric) external
electromagnetic field.
\end{abstract}

\section{Introduction}

\subsection{Equilibrium states}

In the non-relativistic physics two particles can be in equilibrium if the
product of their masses is equal to the product of their charges (in
appropriate units). However, the analogous question in General Relativity is
far from being trivial. Besides the natural mathematical complications, in
General Relativity arise two different types of the \textquotedblright
point\textquotedblright\ centers, namely Kerr - Newman black holes (BH) and
Kerr - Newman naked singularities (NS). Therefore, seeking equilibrium
configurations, one has to consider all three possible types of binary
systems:

\textbullet\ black hole - black hole (BH-BH),

\textbullet\ black hole - naked singularity (BH-NS),

\textbullet\ naked singularity - naked singularity (NS-NS).

Due to the known solitonic generating technique \cite{BV} the construction
of solutions for such binary systems in stationary states do not represents
principal difficulties, but it is quite intricate task to single out from
these solutions the physically reasonable cases because such solutions
possess some features unacceptable from the physical point of view. These
unwanted traits are due to the presence in the solutions the following
exotic peculiarities:

\textbullet\ non-vanishing global NUT parameter,

\textbullet\ closed time-like curves around those parts of the symmetry axis
which are outside the sources,

\textbullet\ angle deficit (or excess) at the points of the symmetry axis,

\textbullet\ non-zero physical magnetic charges of the sources,

\textbullet\ existence of some additional singularities on or off the axis.
\ \ \ \ \ \ \ \ \ \ \ \ \ \ \ \ \ \ \ \ \ \ \ \ \ \ \ \ \ \ \ \ \ \ \ \ \ \
\ \ \ \ \ \ \ \ \ \ \ \ \ \ \ \ \ \ \ \ \ \ \ \ \ \ \ \ \ \ \ \ \ \ \ \ \ \
\ \ \ \ \ \ \ \ \ \ \ \ \ \ \ \ \ \ \ \ \ \ \ \ \ \ \ \ \ \ \ \ \ \ \ \ \ \
\ \ \ \ \ \ \ \ \ \ \ \ \ \ \ \ \ \ \ \ \ \ \ \ \ \ \ \ \ \ \ \ \ \ \ \ \ \
\ \ \ \ \ \ \ \ \ \ \ \ \ \ \ \ \ \ \ \ \ \ \ \ \ \ \ \ \ \ \ \ \ \ \ \ \ \
\ \ \ \ \ \ \ \ \ \ \ \ \ \ \ \ \ \ \ \ \ \ \ \ \ \ \ \ \ \ \ \ \ \ \ \ \ \
\ \ \ \ \ \ \ \ \ \ \ \ \ \ \ \ \ \ \ \ \ \ \ \ \ \ \ \ \ \ \ \ \ \ \ \ \ \
\ \ \ \ \ \ \ \ \ \ \ \ \ \ \ \ \ \ \ \ \ \ \ \ \ \ \ \ \ \ \ \ \ \ \ \ \ \
\ \ \ \ \ \ \ \ \ \ \ \ \ \ \ \ \ \ \ \ \ \ \ \ \ \ \ \ \ \ \ \ \ \ \ \ \ \
\ \ \ \ \ \ \ \ \ \ \ \ \ \ \ \ \ \ \ \ \ \ \ \ \ \ \ \ \ \ \ \ \ \ \ \ \ \
\ \ \ \ \ \ \ \ \ \ \ \ \ \ \ \ \ \ \ \ \ \ \ \ \ \ \ \ \ \ \ \ \ \ \ \ \ \
\ \ \ \ \ \ \ \ \ \ \ \ \ \ \ \ \ \ \ \ \ \ \ \ \ \ \ \ \ \ \ \ \ \ \ \ \ \
\ \ \ \ \ \ \ \ \ \ \ \ \ \ \ \ \ \ \ \ \ \ \ 

To single out solutions without these physically unacceptable properties,
one needs to impose on the parameters of these solitonic families some
constraints which lead to a system of algebraic equations. The problem is
that these algebraic equations (even for the simplest case of two objects)
are extremely complicated and it is difficult to resolve them in order to
show whether they have physically appropriate resolution compatible with the
existence of a positive distance between the sources.

The aforementioned nuisances constitute the real troubles only in the
general case of rotating sources. The comprehensive study of the problem in
static case have been presented in \cite{AB1} where it was constructed the
exact analytic static solution of the Einstein-Maxwell equations for two
charged massive sources separated by the well defined positive distance and
free of aforementioned pathologies. We showed that such static equilibrium
exists only for the BH-NS system and it is impossible to have the physical
equilibrium state for the pairs BH-BH or NS-NS. It is worth mentioning that
these exact results was in full agreement with earlier approximate approach 
\cite{BO1} and numerical experiments \cite{PC}.

After these results, the natural question arose whether the analogous
physical equilibrium exists for two \textit{rotating} sources. It turns out
that for the case BH-NS the answer is affirmative and we were able to
demonstrate \cite{AB2} the exact equilibrium solution of the
Einstein-Maxwell equations for two rotating charged objects one of which is
a Black Hole and another one is a Naked Singularity.

Nevertheless, it is necessary to emphasize that in spite of the known
non-existence of physical equilibrium for the systems BH-BH and NS-NS in
static case, we cannot assert the non-existence of physical equilibrium for
such systems also for rotating objects. May be the rotations can create the
additional forces which can overcome the \textquotedblright no
go\textquotedblright\ result for the static states (although, if so, it
seems to be a little bit strange, because such rotating equilibrium states
would have no limit to the case of vanishing rotations). Thus, for the
binary systems of rotating BH-BH and \ NS-NS the question on the possibility
of physical equilibrium at present remains open.

\subsection{On the influence of the external electromagnetic fields}

Another relevant line of examination the problems of this kind is stability
and existence of the equilibrium states of the binary system of rotating
Black Holes and Naked Singularities immersed in the external electromagnetic
fields. The problem here is to construct solutions describing Black Hole or
Naked Singularity or binary system containing different combinations of such
objects under the influence of the external fields. Of course, the most
interesting cases would be the \textit{exactly} \textit{solvable} models. It
is known that all kind of Black Holes and Naked Singularities and their
binary states are solitons. That's why these states can be described by
exact analytical solutions of the Einstein-Maxwell equations. Is it possible
to immerse these objects into external electromagnetic fields but do not
exceed the bounds of exactly solvable models? The answer is affirmative and
the starting crucial point is that such popular external field as Melvin
magnetic universe \cite{M} indeed can be derived in the framework of the
inverse scattering solitonic generating technique, that is this universe
appear as Einstein-Maxwell soliton ("geon" in the John Wheeler terminology
which nomen have been used by M.A. Melvin in his paper) on the flat
Minkowski space, however, in the form of cosmological non-localizable
configuration\footnote{%
It is worth mentioning that this solution have been obtained ten years
before M.A. Melvin in 1954 by W.B. Bonnor \cite{BO2} [see page 230,
expressions (3.13) in his paper] and has been rediscovered in 1964 by M.A.
Melvin.}. The solitonic nature of the Melvin magnetic universe have been
discovered in 1988 by G.A. Alekseev \cite{A1} in framework of his
generalized inverse scattering technique \cite{A2} applicable for the case
when Einstein-Maxwell fields can be interpreted as external (extending to
infinity) and not in the form of localized perturbations\footnote{%
Such solutions are generated by the meromorphic structure of the dressing
matrix in the complex plane of the spectral parameter but they correspond to
the poles located at infinity of this plane. Mathematically such solutions
represent solitons (because their definition is based only on aforementioned
meromorphicity property without any additional conditions, for example as
localization or finiteness of energy). Although such definition
mathematically corresponds to the notion of solitons in non-gravitational
physics the gravitational solitons in general have many different traits.
For the case of pure gravity these differences have been outlined in
subsection 2.3 of the book \cite{BV}.
\par
Because of the non-local nature of gravitational solitons corresponding to
the pole of dressing matrix at infinity G.A. Alekseev in his papers
preferred called them non-solitonic configurations.}.

The same is true for the Melvin-Harrison universe \cite{H} which generalizes
the Melvin solution by adding an azimuthal twist to the gravitational and
electromagnetic fields. Such a twist appear when together with magnetic also
an electric field is present and these fields (although each of which has no
azimuthal components) are, nevertheless, not parallel producing the flux of
electromagnetic energy in azimuthal direction (that is azimuthal component
of the Poynting vector). This, in turn, produce also non-zero time-azimuthal
component of the metric tensor responsible for the rotation of such
generalized Melvin geon.

Soon after Melvin's publication K.S. Thorne \cite{T} in 1965 investigated
the stability of this solution. The results of his analysis he formulated in
the following clear statements: "It is shown that no radial perturbation can
cause the magnetic field to undergo gravitational collapse to a singularity
or electromagnetic explosion to infinite dispersion. Rather, when
arbitrarily perturbed inside a finite region, the magnetic and gravitational
fields undergo damped, turbulent oscillation until they have radiated away
from the perturbed region all the energy associated with the perturbation.
Then they settle down into Melvin's unperturbed, static configuration". Such
conclusion suggests that the Melvin solution in some finite region of space
can have a reasonable physical application.

The first step in this direction was made in 1974 by R.M. Wald \cite{W} who
(on the base of A. Papapetrou analysis of some properties of Killing vectors
in case of axial symmetry \cite{P}) constructed the exact solution for the 
\textit{test} electromagnetic field in the fixed external gravitational
field produced by the Kerr black hole. Far from black hole the Wald
space-time (in the linear approximation with respect to the electromagnetic
potentials) tends to the Melvin magnetic universe.

This field of research have been essentially promoted in 1976 by F.J. Ernst
and W.J. Wild \cite{EW} who constructed the \textit{exact }solution of
Einstein-Maxwell equation describing the Kerr-Newman black hole placed in
the Melvin magnetic universe. This solution have been obtained by different
technique with respect to our inverse scattering method but, of course, our
"solitonic approach" for this particular case should give the same result.

\subsection{Method in general}

The Melvin magnetic field configuration can be generalized further and
corresponding generalization we can call Melvin Class of solutions of the
Einstein-Maxwell equations. As we already mentioned, the general theory and
the method how to construct such non-localazible solutions in the framework
of the inverse scattering technique have been developed by G.A. Alekseev 
\cite{A1, A2}. Then it is possible to construct a set of exact solutions of
the Einstein-Maxwell equations representing black holes (BH), naked
singularities (NS) or their combinations immersed into the Melvin Class
external electromagnetic fields.

To do this one need first of all solutions generating technique for the
elementary processes $(BH)_{0}\rightarrow (BH)_{deformed}+External$ $Field$
or $(NS)_{0}\rightarrow (NS)_{deformed}+External$ $Field$ where $(BH)_{0}$
and $(NS)_{0}$ are initial background configurations. In general, solutions
generating technique permits to solve the matrix linear eigenvalue problem%
\begin{equation}
\partial \mathbf{\varphi }\left( x,w\right) /\partial x^{\mu }=\Lambda _{\mu
}{}^{\nu }\left( x,w\right) \mathbf{U}_{\nu }\left( x\right) \mathbf{\varphi 
}\left( x,w\right) \text{ },  \label{V1}
\end{equation}%
the integrability conditions of which coincide with original
Einstein-Maxwell equations. Here the components $\Lambda _{\mu }{}^{\nu
}\left( x,w\right) $ are given (solution independent) functions of $x^{\mu }$%
, $w$ and $w$ is a complex spectral parameter independent on the coordinates 
$x^{\mu }$. The seed (background) solution for the metric and
electromagnetic potentials together with seed matrices $\mathbf{U}_{\nu
}^{(0)}$ and $\mathbf{\varphi }^{(0)}$ are known. Then perturbed matrix $%
\mathbf{\varphi }$ can be found by the dressing procedure $\mathbf{\varphi }=%
\mathbf{\chi \varphi }^{(0)}$ solving the corresponding spectral equation
for $\mathbf{\chi }$ for\ each type of its analytical structure in $w$%
-plane. The standard localized solitonic perturbations correspond to the
meromorphic matrix $\mathbf{\chi }$ tending to identity at $w\rightarrow
\infty $ and having the finite number of poles in the finite region of
complex $w$\textit{-}plane. The new point is that Melvin Class of external
fields is generated also by meromorphic structure of dressing matrix $\chi $
but with poles at infinity $w\rightarrow \infty $. In general this
corresponds to the rational function (with respect to the spectral parameter 
$w$) analytical structure of the dressing matrix:%
\begin{equation}
\mathbf{\chi =}\sum_{k=1}^{k=n}\frac{\mathbf{R}_{k}\left( x\right) }{w-w_{k}}%
+\mathbf{A}^{\left( 0\right) }\left( x\right) +w\mathbf{A}^{\left( 1\right)
}\left( x\right) +w^{2}\mathbf{A}^{\left( 2\right) }\left( x\right)
+...+w^{m}\mathbf{A}^{\left( m\right) }\left( x\right) \text{ }.  \label{V2}
\end{equation}

Here the first sum of simple poles at finite region of the $w$-plane
generates the set of $n$ localized solitons (the multiple poles localizable
solitons also can be obtained by the poles fusion procedure \cite{BV}, like
it was done for Tomimatsu-Sato multi-solitonic solution). In stationary case
these are a mixture of $n$ interacting black holes and naked singularities.
The multiple pole at infinity generates external electromagnetic fields into
which this mixture is immersed. Up to now we show in details how to
construct exact solution of the Einstein-Maxwell equations for the simplest
case of the such dressing matrix (one simple pole in finite region and one
simple pole at infinity): 
\begin{equation}
\mathbf{\chi =}\frac{\mathbf{R}_{1}\left( x\right) }{w-w_{1}}+\mathbf{A}%
^{\left( 0\right) }\left( x\right) +w\mathbf{A}^{\left( 1\right) }\left(
x\right) \text{ }.  \label{V3}
\end{equation}

Depending on the free arbitrary parameters the external fields contain we
have solutions of the three types:

1. Kerr-Newman object placed into external electromagnetic field which has
original Melvin pure magnetic structure at spatial infinity (this is the
solution which has been obtained in \cite{EW}).

2. The same but with twisting Melvin-Harrison gravitational and
electromagnetic fields at spatial infinity \cite{H}.

3. The set of solutions corresponding to the general Melvin Class
asymptotics at spatial infinity. Some subset of these solutions have the new
exotic peculiarities outside the symmetry axis (most interesting \ are
ergoregions of the toroidal structure), see discussion in \cite{A1}.

Case 3 are new but a number of these solutions need a remedy to be made free
from some unphysical phenomenons. We think that this can be possible by
adding multipole poles at $w\rightarrow \infty $.

\section{Einstein-Maxwell equations}

The task of the inverse scattering method for the interacting gravitational
and electromagnetic fields is to find a new solution of Einstein-Maxwell
equations if some background (seed) of these equations is given. The method
represents the generalization of the pure gravity inverse scattering
technique \cite{BZ1,BZ2} and has been developed by G.A. Alekseev in \cite{A3}
which technique he exposed in details in \cite{A1}. We described Alekseev
technique (in slightly different notations) in chapter "Einstein-Maxwell
fields" of the book \cite{BV} and in the present paper we follow definitions
and notations from this book.

The outline of the method is as follows. For the integrable ansatz of the
Einstein--Maxwell equations the metrics is:%
\begin{equation}
ds^{2}=f\eta _{\mu \nu }dx^{\mu }dx^{\nu }+g_{ab}dx^{a}dx^{b},  \label{I1}
\end{equation}%
where the Greek indices as well as small Latin indices take only two values.
The metric coefficients $f$ and $g_{ab}$ are functions only on two Greek
coordinates $x^{\mu }$. In case of stationary fields with axial symmetry the
variables $x^{\mu }$ correspond to the standard pair of cylindrical
coordinates $(\rho ,z)$ (that is radial distance from the axis and
coordinate along this axis) and Latin variables $x^{a}$ stand for time and
azimuthal angle. In case of non-stationary fields the pair of $x^{\mu }$
signify time and one active spatial coordinate while the variables $x^{a}$
correspond to the two residual (mute) spatial variables from which the
fields do not depend. Accordingly, the metric components $\eta _{\mu \nu }$
take the form:%
\begin{equation}
\eta _{\mu \nu }=\left( 
\begin{array}{cc}
-e & 0 \\ 
0 & 1%
\end{array}%
\right) \text{ },  \label{I2}
\end{equation}%
where $e=-1$ and $e=1$ for the stationary and non-stationary cases,
respectively.

From now on we attribute to the small Latin indices numerical values $1,2$
(as for the Greek indices one also can keep in mind some values for them but
nowhere in the paper there is a necessity to use for these indexes any
concrete numerics). With this convention the interval (\ref{I1}) is:%
\begin{equation}
ds^{2}=f\eta _{\mu \nu }dx^{\mu }dx^{\nu }+g_{11}\left( dx^{1}\right)
^{2}+g_{22}\left( dx^{2}\right) ^{2}+2g_{12}dx^{1}dx^{2},  \label{Dop1}
\end{equation}%
The metric coefficients $g_{ab}$ are real and symmetric ($g_{21}=g_{12}$)
and for the determinant $g_{11}g_{22}-g_{12}^{2}$ we adopt the notation:%
\begin{equation}
g_{11}g_{22}-g_{12}^{2}=e\alpha ^{2}.  \label{I3}
\end{equation}

In order to make the integrable ansatz compatible with the metric of this
type one should assume that only components $A_{a}$ of the electromagnetic
potentials are non zero and they depend only on Greek coordinates $x^{\mu }$%
, that is:%
\begin{equation}
A_{\mu }=0\text{ },\text{ }A_{a,b}=0\text{ },\text{ }A_{a,\mu }\neq 0.
\label{I4}
\end{equation}%
Correspondingly, the nonvanishing covariant and contravariant components of
the electromagnetic field tensor are: 
\begin{equation}
F_{\mu a}=A_{a,\mu }\text{ },\text{ }F^{\mu a}=f^{-1}\eta ^{\mu \nu
}g^{ab}A_{b,\nu }\text{ }.  \label{I5}
\end{equation}%
Here, and in the sequel, $\eta ^{\mu \nu }$ and $g^{ab}$ are the metric
components inverse to $\eta _{\mu \nu }$ and $g_{ab}$ respectively (that is $%
\eta ^{\mu \lambda }\eta _{\lambda \nu }=\delta _{\nu }^{\mu },$ $%
g^{ac}g_{cb}=\delta _{b}^{a}$) and by comma we denote the simple partial
derivatives.

The Einstein-Maxwell equations for the metric coefficients $g_{ab}$ and
electromagnetic potentials $A_{a}$ do not contain the scale factor $f$ and
represent the following closed self-consistent system:%
\begin{equation}
\eta ^{\mu \nu }\alpha ^{-1}(\alpha g^{bc}g_{ac,\mu })_{,\nu }=-4\eta ^{\mu
\nu }g^{bc}A_{a,\mu }A_{c,\nu }+2\delta _{a}^{b}\eta ^{\mu \nu
}g^{cd}A_{c,\mu }A_{d,\nu }\text{ },  \label{I6}
\end{equation}%
\begin{equation}
\eta ^{\mu \nu }(\alpha g^{ab}A_{b,\mu })_{,\nu }=0\text{ }.  \label{I7}
\end{equation}%
Contracting indices $a$ and $b$ in the first of these equations we get the
useful consequence:%
\begin{equation}
\eta ^{\mu \nu }\alpha _{,\mu \nu }=0\text{ }.  \label{I8}
\end{equation}%
If $\alpha $ is some given solution of this equation, the second
functionally independent solution we denote by $\beta $ and it can be found
(up to an arbitrary additive constant) from the relation 
\begin{equation}
\beta _{,\mu }=-e\eta _{\mu \rho }\epsilon ^{\rho \sigma }\alpha _{,\sigma }%
\text{ }.\text{ }  \label{I9}
\end{equation}%
Here and in what follows we use the \textit{Euclidean} Kroneker symbols in
Greek indices as%
\begin{equation}
\epsilon ^{\mu \nu }=\epsilon _{\mu \nu }=\left( 
\begin{array}{cc}
0 & 1 \\ 
-1 & 0%
\end{array}%
\right) \text{ },  \label{I10}
\end{equation}%
and the same (which we will need later) for the Latin indices:%
\begin{equation}
\epsilon ^{ab}=\epsilon _{ab}=\left( 
\begin{array}{cc}
0 & 1 \\ 
-1 & 0%
\end{array}%
\right) \text{ }.  \label{I11}
\end{equation}

Alongside with $A_{a}$ we introduce an auxiliary electromagnetic potentials $%
B_{a}$ which play an intermediary role and not manifest itself in the final
results, however, which is very helpful for technical side of the inverse
scattering machinery. This are defined by the following differential
relation:%
\begin{equation}
B_{a,\mu }=-\alpha ^{-1}\eta _{\mu \nu }\epsilon ^{\nu \lambda
}g_{ab}\epsilon ^{bc}A_{c,\lambda }\text{ }.  \label{I12}
\end{equation}%
It is easy to verify that the integrability condition for this equation, $%
\epsilon ^{\mu \nu }B_{a,\mu \nu }=0,$ coincides with the Maxwell equation (%
\ref{I7}). Relation (\ref{I12}) can also be written in its inverse form:%
\begin{equation}
A_{a,\mu }=\alpha ^{-1}\eta _{\mu \nu }\epsilon ^{\nu \lambda
}g_{ab}\epsilon ^{bc}B_{c,\lambda }\text{ }.  \label{I13}
\end{equation}%
Now we combine $A_{a}$ and $B_{a}$ into a single complex electromagnetic
potential $\Phi _{a}$ defined by%
\begin{equation}
\Phi _{a}=A_{a}+iB_{a}\text{ },  \label{I14}
\end{equation}%
then (\ref{I12})--(\ref{I13}) are, respectively, the imaginary and real
parts of the following equation for $\Phi _{a}$:%
\begin{equation}
\Phi _{a,\mu }=-i\alpha ^{-1}\eta _{\mu \nu }\epsilon ^{\nu \lambda
}g_{ab}\epsilon ^{bc}\Phi _{c,\lambda }\text{ },  \label{I15}
\end{equation}%
from which the Maxwell equations for $\Phi _{a}$ trivially results:%
\begin{equation}
\eta ^{\mu \nu }(\alpha g^{ab}\Phi _{b,\mu })_{,\nu }=0\text{ }.  \label{I16}
\end{equation}%
By the direct calculation one can show that Einstein equations (\ref{I6})
can be written as

\begin{equation}
\eta ^{\mu \nu }\alpha ^{-1}(\alpha g^{bc}g_{ac,\mu })_{,\nu }=-2g^{bc}\eta
^{\mu \nu }\bar{\Phi}_{a,\mu }\Phi _{c,\nu }\text{ },  \label{I17}
\end{equation}%
where (and in what follows) the line over a letter signify complex
conjugation. Due to relation (\ref{I12}) the right hand side of this
equation is real and coincides with the right hand side of (\ref{I6}). It is
thus clear that any solution $g_{ab\text{ }}$and $A_{a}=\func{Re}\Phi _{a}$
of (\ref{I14})--(\ref{I17}) is also a solution of the Einstein--Maxwell
equations (\ref{I6})--(\ref{I7}). Note the auxiliary role played by the
potential $B_{a}$.

Einstein-Maxwell equations can be represented in more compact form using
three-dimensional matrices (we designate them by the bold letters). Let's
use the capital Latin indexes $A,B,...$ to enumerate the components of these
matrices. Each such index take values $1,2,3$ and fuse previously used small
Latin indexes with index $3$ which enumerate the third rows and columns of
the three-dimensional matrices.

We note for definiteness that in any matrix components of the types $\left( 
\mathbf{M}\right) ^{AB},$ $\left( \mathbf{M}\right) _{AB}$ $,$ $\left( 
\mathbf{M}\right) ^{A}{}_{B}$ or $\left( \mathbf{M}\right) _{A}{}^{B}$ the
first index, independent of its up or \ down position, will enumerate the
rows, and the second index - columns.

The basic objects for such 3-dimensional representation of the theory are 
\textit{Hermitian} matrix $\mathbf{G}$ (with components $\left( \mathbf{G}%
\right) ^{AB}$):%
\begin{equation}
\mathbf{G=}\left( 
\begin{array}{ccc}
4g_{22}+4\Phi _{2}\bar{\Phi}_{2} & -4g_{21}-4\Phi _{2}\bar{\Phi}_{1} & 2\Phi
_{2} \\ 
-4g_{12}-4\Phi _{1}\bar{\Phi}_{2} & 4g_{11}+4\Phi _{1}\bar{\Phi}_{1} & 
-2\Phi _{1} \\ 
2\bar{\Phi}_{2} & -2\bar{\Phi}_{1} & 1%
\end{array}%
\right) \text{ },  \label{I20}
\end{equation}%
and its inverse $\mathbf{G}^{-1}$ (having components $\left( \mathbf{G}%
^{-1}\right) _{AB}$):%
\begin{equation}
\mathbf{G}^{-1}=\left( 
\begin{array}{ccc}
\frac{1}{4}g^{22} & -\frac{1}{4}g^{21} & -\frac{1}{2}\Phi ^{2} \\ 
-\frac{1}{4}g^{12} & \frac{1}{4}g^{11} & \frac{1}{2}\Phi ^{1} \\ 
-\frac{1}{2}\bar{\Phi}^{2} & \frac{1}{2}\bar{\Phi}^{1} & 1+\bar{\Phi}%
^{a}\Phi _{a}%
\end{array}%
\right) \text{ },  \label{I21}
\end{equation}%
where $\Phi ^{a}$ is defined as:%
\begin{equation}
\Phi ^{a}=g^{ab}\Phi _{b}\text{ }.  \label{Dop2}
\end{equation}%
From (\ref{I20}) follows the determinant of matrix $\mathbf{G}$:%
\begin{equation}
\det \mathbf{G=}16\left( g_{11}g_{22}-g_{12}^{2}\right) \text{ }.
\label{Dop4}
\end{equation}

Now the Einstein-Maxwell equations (\ref{I16})-(\ref{I17}) together with
additional condition (\ref{I15}) acquire the three-dimensional matrix form: 
\begin{equation}
\eta ^{\mu \nu }(\alpha \mathbf{YG}_{,\mu }\mathbf{G}^{-1})_{,\nu }=0\text{ }
\label{I18}
\end{equation}%
where the matrix $\mathbf{Y}$ (with components $\left( \mathbf{Y}\right)
^{A}{}_{B}$) are defined as%
\begin{equation}
\mathbf{Y=}\left( 
\begin{array}{ccc}
1 & 0 & 2\Phi _{2} \\ 
0 & 1 & -2\Phi _{1} \\ 
0 & 0 & 2%
\end{array}%
\right)  \label{Dop3}
\end{equation}

Substitution of (\ref{I20}), (\ref{I21}) and (\ref{Dop3}) into (\ref{I18})
gives exactly (\ref{I15})-(\ref{I17}) and nothing else.

After matrices $\mathbf{G}$ and $\mathbf{Y}$ become known the metric
coefficient $f$ can be found by quadratures from the following equation:%
\begin{gather}
(\ln f)_{,\mu }=(\ln D)_{,\mu }-(\ln \alpha )_{,\mu }  \label{I22} \\
+\frac{(\ln \alpha )_{,\lambda }}{4D}\left[ 2\eta ^{\lambda \sigma }Tr\left( 
\mathbf{K}_{\mu }\mathbf{K}_{\sigma }\right) -\delta _{\mu }^{\lambda }\eta
^{\rho \sigma }Tr\left( \mathbf{K}_{\rho }\mathbf{K}_{\sigma }\right) \right]
\text{ },  \notag
\end{gather}%
where%
\begin{equation}
D=\eta ^{\mu \nu }\alpha _{,\mu }\alpha _{,\nu }\text{ },\text{ }\mathbf{K}%
_{\mu }=\alpha \mathbf{YG}_{,\mu }\mathbf{G}^{-1}\text{ }.  \label{I23}
\end{equation}%
The integrability condition for equation (\ref{I22}), $\epsilon ^{\mu \nu
}\left( \ln f\right) _{,\mu \nu }=0,$ satisfy automatically if the previous
equations for the metric coefficients $g_{ab}$ and potentials $\Phi _{a}$
are satisfied. The complete Einstein-Maxwell system contain also one
additional differential equation of the second order for the metric
coefficient $f$. However, it is the direct consequence of the relations (\ref%
{I22})-(\ref{I23}) and of all previous equations for $g_{ab}$ and $\Phi _{a}$%
, then there is no necessity to pay attention to it.

\section{Construction of solutions}

We see that any solution of Einstein-Maxwell equations is encoded in matrix $%
\mathbf{G}$ which not only satisfy equation (\ref{I18}) but also has special
structure (\ref{I20})-(\ref{I21}) [note that a solution of equations (\ref%
{I18})-(\ref{Dop3}) in general has no bound to give matrix $\mathbf{G}$
namely in the special form (\ref{I20})-(\ref{I21})]. Correspondingly we need
to construct a three-dimensional linear spectral problem the
self-consistency conditions of which provide for both of these properties.
This task is resolved by the following spectral equation for
three-dimensional matrix $\mathbf{\varphi }$ (with components $\left( 
\mathbf{\varphi }\right) _{AB}$):%
\begin{equation}
\mathbf{\varphi }_{,\mu }=\frac{\left( w+\beta \right) \delta _{\mu }^{\nu
}+e\alpha \eta _{\mu \rho }\epsilon ^{\rho \nu }}{2i\left[ \left( w+\beta
\right) ^{2}-e\alpha ^{2}\right] }\mathbf{U}_{\nu }\mathbf{\varphi }\text{ },
\label{I24}
\end{equation}%
where $w$ is free complex spectral parameter independent on coordinates $%
x^{\mu }$. Matrix $\mathbf{\varphi }$ is function on all three variables $%
x^{\mu }$ and $w,$ but matrices $\mathbf{U}_{\nu }$ (with components $\left( 
\mathbf{U}_{\nu }\right) _{A}{}^{B}$)\textbf{\ } depend only on two
coordinates $x^{\mu }$ and are of the form: 
\begin{equation}
\mathbf{U}_{\nu }=ie\alpha \eta _{\nu \rho }\epsilon ^{\rho \sigma }\mathbf{G%
}^{-1}\mathbf{G}_{,\sigma }\mathbf{Y}^{\dagger }+4e(\alpha ^{2}\mathbf{G}%
^{-1})_{,\nu }\mathbf{\Omega }\text{ },  \label{I25}
\end{equation}%
where matrix $\mathbf{G}$ and $\mathbf{Y}$ have been introduced by
expressions (\ref{I20}),(\ref{I21}) and (\ref{Dop3}). Matrix $\mathbf{\Omega 
}$ (with components $\left( \mathbf{\Omega }\right) ^{AB}$) is defined as\ 
\begin{equation}
\mathbf{\Omega }=\left( 
\begin{array}{ccc}
0 & 1 & 0 \\ 
-1 & 0 & 0 \\ 
0 & 0 & 0%
\end{array}%
\right) .  \label{I26}
\end{equation}%
Direct calculations show that the integrability conditions for the equation (%
\ref{I24}) (that is $\mathbf{\varphi }_{,\mu \nu }=\mathbf{\varphi }_{,\nu
\mu }$) coincide with equation (\ref{I18}) and, due to the structure (\ref%
{I20}),(\ref{I21}) and (\ref{Dop3}) of matrices $\mathbf{G}$ and $\mathbf{Y}$%
, gives Einstein-Maxwell equations (\ref{I15})-(\ref{I17}).

Then it is clear that any solution for $\mathbf{\varphi }$ automatically
contains solution for Einstein-Maxwell equations encoded in matrix $\mathbf{G%
}$. Now the problem is how to pull out matrix $\mathbf{G}$ from $\mathbf{%
\varphi }$.

The inverse scattering dressing procedure implies that we need to know some
background solution $g_{ab}^{\left( 0\right) },\Phi _{a}^{\left( 0\right) }$
of the Einstein-Maxwell equations for metric coefficients and
electromagnetic potentials from which we can get from (\ref{I20})-(\ref{I21}%
) the background matrix $\mathbf{G}^{\left( 0\right) }$ as well as its
inverse. Then from relation (\ref{I25}) follows the background matrices $%
\mathbf{U}_{\mu }^{\left( 0\right) }$ and from differential equation (\ref%
{I24}) we can find the background spectral matrix $\mathbf{\varphi }^{\left(
0\right) }$ (the calculation of $\mathbf{\varphi }^{\left( 0\right) }$ is
the only point in the procedure which need a real integration of some
differential equations, all other steps of the method consist either of pure
algebraic manipulation or calculation the quadratures). The dressing
procedure for construction a new solutions from the given background one
means that the new spectral matrix $\mathbf{\varphi }$ can be represented as
"dressed" value of its seed solution $\mathbf{\varphi }^{\left( 0\right) }$,
that is: 
\begin{equation}
\mathbf{\varphi }=\mathbf{\chi \varphi }^{\left( 0\right) },  \label{I27}
\end{equation}%
where matrix $\mathbf{\chi }$ (with components $\left( \mathbf{\chi }\right)
_{A}{}^{B}$) play a role of relative exact perturbation of the background
and depends on all three variables $x^{\mu }$ and $w$. Equation for matrix $%
\mathbf{\chi }$ following from (\ref{I24}) [taking into account that $%
\mathbf{\varphi }^{\left( 0\right) }$ is a solution of (\ref{I24})] is:%
\begin{equation}
\mathbf{\chi }_{,\mu }=\frac{\left( w+\beta \right) \delta _{\mu }^{\nu
}+e\alpha \eta _{\mu \rho }\epsilon ^{\rho \nu }}{2i\left[ \left( w+\beta
\right) ^{2}-e\alpha ^{2}\right] }\left( \mathbf{U}_{\nu }\mathbf{\chi -\chi
U}_{\nu }^{\left( 0\right) }\right) \text{ }.  \label{I28}
\end{equation}

The analysis shows that due to equations (\ref{I24}),(\ref{I27}),(\ref{I28})
and special structure (\ref{I20})-(\ref{I21}) of matrix $\mathbf{G}$ we have
the following important algebraic relation:%
\begin{equation}
\mathbf{G+}4i\left( w+\beta \right) \mathbf{\Omega }=\left( \mathbf{\chi }%
^{-1}\right) ^{\dagger }\left[ \mathbf{G}^{\left( 0\right) }+4i\left(
w+\beta \right) \mathbf{\Omega }\right] \mathbf{\chi }^{-1}\text{ },
\label{I29}
\end{equation}%
which is satisfied everywhere in the complex plane of parameter $w$. Here
appeared the Hermitian conjugation of the matrix $\mathbf{\chi }^{-1}$(with
components $\left( \mathbf{\chi }^{-1}\right) _{A}{}^{B}$) which depends on $%
w$ and the sense of such object need to be explained. The definition of the
Hermitian conjugation of matrix functions depending on the real coordinates $%
x^{\mu }$ and complex parameter $w$ is the following: to obtain the
Hermitian conjugation $\mathbf{M}^{\dagger }$ as a function of $x^{\mu }$
and $w$ of any matrix $\mathbf{M}$ one should first calculate the value of
the matrix $\mathbf{M}$ at the complex conjugate point $\bar{w}$, i.e. the
value $\mathbf{M(}x^{\mu },\bar{w})$, and then take the usual Hermitian
conjugate of this value.

Namely from equation (\ref{I29}) follows matrix $\mathbf{G}$ (and,
consequently, metric and electromagnetic potentials) in terms of the
background solution if also the dressing matrix $\mathbf{\chi }$ can be
found in terms of the background quantities. However, precisely this
miraculous possibility is provided by the inverse scattering method in case
of the meromorphic structure of the dressing matrices $\mathbf{\chi }$ and $%
\mathbf{\chi }^{-1}$ with respect to the parameter $w$. In this way the
final expressions for the metric and electromagnetic potentials can be
obtained in exact analytical form.

\section{Alekseev's extension for the dressing matrix}

In the book \cite{BV} we considered only the Einstein-Maxwell solitonic
solutions when both dressing matrix $\mathbf{\chi }$ and its inverse $%
\mathbf{\chi }^{-1}$ have simple poles at the \textit{finite} points of the
complex plane of the spectral parameter $w$ and also under condition that
all these points are different from each other. Here, following the Alekseev
papers \cite{A1, A2}, we describe the generalization of this technique for
the case when dressing matrices $\mathbf{\chi }$ and $\mathbf{\chi }^{-1}$
both have one simple pole at infinity of the complex plane of spectral
parameter. In this case $\mathbf{\chi }$ and $\mathbf{\chi }^{-1}$ are of
the form:%
\begin{equation}
\mathbf{\chi =A}w\mathbf{+P}\text{ }\mathbf{,}\text{ }\mathbf{\chi }^{-1}%
\mathbf{=B}w\mathbf{+Q}\text{ }\mathbf{,}  \label{A1}
\end{equation}%
where matrices $\mathbf{A},$\textbf{\ }$\mathbf{P},$\textbf{\ }$\mathbf{B},$%
\textbf{\ }$\mathbf{Q}$\textbf{\ }(with components $\left( \mathbf{A}\right)
_{A}{}^{B},\left( \mathbf{P}\right) _{A}{}^{B},\left( \mathbf{B}\right)
_{A}{}^{B},\left( \mathbf{Q}\right) _{A}{}^{B}$\textbf{) }do not depend on
the parameter $w$ and matrix $\mathbf{P}$ has an inverse $\mathbf{P}^{-1}$
(with components $\left( \mathbf{P}^{-1}\right) _{A}{}^{B}$).

Complete system where from all matrices can be found are two evident
relations $\mathbf{\chi \chi }^{-1}=\mathbf{I}$, $\mathbf{\chi }^{-1}\mathbf{%
\chi }=\mathbf{I}$ and equations (\ref{I28}), (\ref{I29}). Each of these can
be represented as zero condition for some polynomial (cubic or second order)
with respect to the parameter $w$. The vanishing conditions for all
coefficients of the second order $w$-polynomials $\mathbf{\chi \chi }^{-1}-%
\mathbf{I}$ and $\mathbf{\chi }^{-1}\mathbf{\chi }-\mathbf{I}$ give matrices 
$\mathbf{B},$\textbf{\ }$\mathbf{Q}$ as: 
\begin{equation}
\mathbf{Q}=\mathbf{P}^{-1},\text{ \ }\mathbf{B}=-\mathbf{P}^{-1}\mathbf{AP}%
^{-1},  \label{A2}
\end{equation}%
and the special requirement for matrix $\mathbf{A}$:%
\begin{equation}
\mathbf{AP}^{-1}\mathbf{A}=0\text{ }.  \label{A3}
\end{equation}%
Then from (\ref{I28}) and (\ref{I29}) first of all follows two additional
demands for the matrix $\mathbf{A}$:%
\begin{equation}
\mathbf{A}_{,\mu }=0\text{ },\text{ }  \label{A4}
\end{equation}
\begin{equation}
\mathbf{A}^{\dagger }\mathbf{\Omega A=0}  \label{A5}
\end{equation}%
and basic expressions for matrices $\mathbf{G}$ and $\mathbf{U}_{\mu }$:%
\begin{equation}
\mathbf{G+}4i\beta \mathbf{\Omega =}\left( \mathbf{P}^{\dagger }\right)
^{-1}\left( \mathbf{G}^{\left( 0\right) }\mathbf{+}4i\beta \mathbf{\Omega }%
\right) \mathbf{P}^{-1}\text{ },  \label{A6}
\end{equation}%
\begin{equation}
\mathbf{U}_{\mu }=2i(\beta \delta _{\mu }^{\nu }-e\alpha \eta _{\mu \rho
}\epsilon ^{\rho \nu })\mathbf{P}_{,\nu }\mathbf{P}^{-1}+\mathbf{PU}_{\mu
}^{(0)}\mathbf{P}^{-1}.  \label{A7}
\end{equation}%
Using equations (\ref{A3})-(\ref{A7}) all the rest requirements following
from (\ref{I28}) and (\ref{I29}) can be written in the form:

\begin{equation}
2i\mathbf{P}^{-1}\mathbf{P}_{,\mu }-2i(\beta \delta _{\mu }^{\nu }-e\alpha
\eta _{\mu \rho }\epsilon ^{\rho \nu })\mathbf{P}^{-1}\mathbf{P}_{,\nu }%
\mathbf{P}^{-1}\mathbf{A}=\mathbf{U}_{\mu }^{\left( 0\right) }\mathbf{P}^{-1}%
\mathbf{A}-\mathbf{P}^{-1}\mathbf{AU}_{\mu }^{(0)}\text{ },  \label{A8}
\end{equation}%
\begin{equation}
\left( \mathbf{P}^{-1}\mathbf{A}\right) ^{\dagger }\left( \mathbf{G}^{\left(
0\right) }\mathbf{+}4i\beta \mathbf{\Omega }\right) \mathbf{P}^{-1}\mathbf{A+%
}4i\left( \mathbf{P}^{\dagger }\mathbf{\Omega A+A}^{\dagger }\mathbf{\Omega P%
}\right) \mathbf{=0}\text{ },  \label{A9}
\end{equation}%
\begin{equation}
\left( \mathbf{G}^{\left( 0\right) }\mathbf{+}4i\beta \mathbf{\Omega }%
\right) \mathbf{P}^{-1}\mathbf{A+}\left( \mathbf{P}^{-1}\mathbf{A}\right)
^{\dagger }\left( \mathbf{G}^{\left( 0\right) }\mathbf{+}4i\beta \mathbf{%
\Omega }\right) \mathbf{+}4i\mathbf{P}^{\dagger }\mathbf{\Omega P=}4i\mathbf{%
\Omega }\text{ }.  \label{A10}
\end{equation}

The six equations (\ref{A3})-(\ref{A5}) and (\ref{A8})-(\ref{A10}) represent
the complete self-consistent system from which matrices $\mathbf{A}$ and $%
\mathbf{P}$ can be found in terms of the background matrices $\mathbf{G}%
^{\left( 0\right) },$ $\mathbf{U}_{\mu }^{\left( 0\right) }$ and given
(chosen from the outset) functions $\alpha ,\beta $. Remarkably enough this
system can be resolved exactly \cite{A1} and it has a unique solution up to
the three arbitrary real constants, which represent free parameters of the
dressed solution generated by the pole at infinity of the $w$-plane. After
we obtain in this way matrix $\mathbf{P}$ the resulting solutions of
interest for $\mathbf{G}$ and $\mathbf{U}_{\mu }$ follow from equations (\ref%
{A6}) and (\ref{A7}). The results show that $\mathbf{A}$ is the degenerate
matrix of the form:%
\begin{equation}
\mathbf{A=}\left( 
\begin{array}{ccc}
8ik_{1}k_{2} & -8ik_{1}^{2} & 0 \\ 
0 & 0 & 0 \\ 
0 & 0 & 0%
\end{array}%
\right) \text{ },  \label{A11}
\end{equation}%
where $k_{1}$ and $k_{2}$ are two arbitrary real constants and matrix $%
\mathbf{P}$ has the following structure:%
\begin{equation}
\mathbf{P=}\left( 
\begin{array}{ccc}
2k_{3}N-k_{2}\kappa \left( L+2QN\right) -\kappa ^{-1} & k_{1}\kappa \left(
L+2QN\right) & -2k_{1}N \\ 
-k_{2}k_{1}^{-1}\kappa & \kappa & 0 \\ 
k_{3}k_{1}^{-1}-k_{2}k_{1}^{-1}\kappa Q & \kappa Q & -1%
\end{array}%
\right) \text{ }.  \label{A12}
\end{equation}%
Matrix $\mathbf{P}^{-1}$ inverse to $\mathbf{P}$ is:%
\begin{equation}
\mathbf{P}^{-1}=\left( 
\begin{array}{ccc}
-\kappa & k_{1}\kappa L & 2k_{1}\kappa N \\ 
-k_{2}k_{1}^{-1}\kappa & k_{2}\kappa L+\kappa ^{-1} & 2k_{2}\kappa N \\ 
-k_{3}k_{1}^{-1}\kappa & k_{3}\kappa L+Q & 2k_{3}\kappa N-1%
\end{array}%
\right) \text{ }.  \label{A13}
\end{equation}%
Here $k_{3}$ is the third arbitrary constant parameter which is complex
number of the unit modulus:%
\begin{equation}
k_{3}=e^{i\gamma }\text{ },  \label{A14}
\end{equation}%
where $\gamma $ is an arbitrary real constant. It is easy to show that 
\begin{equation}
\det \mathbf{P}=1\text{ },  \label{Dop5}
\end{equation}

The three constants $k_{1},k_{2}$ and $k_{3}$ we organize in form of
3-dimensional object $k_{A\text{ }}$:%
\begin{equation}
k_{A}=\left( k_{1},k_{2},k_{3}\right) \text{ }.\text{ }  \label{A15}
\end{equation}%
Matrix $\mathbf{P}$ contains four complex valued functions $\kappa ,$ $N,$ $%
Q,$ $L$ depending on the real Greek coordinates $x^{\mu }$. These functions
can be found in terms of the arbitrary constants $k_{A}$, background
matrices $\mathbf{G}^{\left( 0\right) },$ $\mathbf{U}_{\mu }^{\left(
0\right) }$ and given functions $\alpha $ and $\beta $.

Real part of function $\kappa ^{-1}$ can be obtained explicitly:%
\begin{equation}
\kappa ^{-1}+\bar{\kappa}^{-1}=2\bar{k}_{A}k_{B}(\mathbf{G}^{\left( 0\right)
})^{AB}\text{ },  \label{A16}
\end{equation}%
and its imaginary part should be calculated from the quadrature%
\begin{equation}
(\kappa ^{-1}-\bar{\kappa}^{-1})_{,\mu }=-4\bar{k}_{C}k_{B}\left[ (\mathbf{%
\Omega })^{CA}\left( \mathbf{U}_{\mu }^{\left( 0\right) }\right) _{A}{}^{B}-(%
\mathbf{\Omega })^{BA}\left( \mathbf{\bar{U}}_{\mu }^{\left( 0\right)
}\right) _{A}{}^{C}\right] \text{ }.  \label{A17}
\end{equation}%
It is worth noting that from the last two equations and from equation (\ref%
{I25}) applying to the background solution can be derived following useful
consequence:%
\begin{equation}
(\kappa ^{-1})_{,\mu }=-4k_{B}\bar{k}_{C}(\mathbf{\Omega })^{CA}\left( 
\mathbf{U}_{\mu }^{\left( 0\right) }\right) _{A}{}^{B}  \label{Dop6}
\end{equation}

The quantity $N$ has simple explicit expression:%
\begin{equation}
N=\bar{k}_{A}\left( \mathbf{G}^{\left( 0\right) }\right) ^{A3}\text{ },
\label{A18}
\end{equation}%
and function $Q$ follows from quadrature:%
\begin{equation}
Q_{,\mu }=4k_{C}\left[ k_{3}\left( \mathbf{U}_{\mu }^{\left( 0\right)
}\right) _{1}{}^{C}-k_{1}\left( \mathbf{U}_{\mu }^{\left( 0\right) }\right)
_{3}{}^{C}\right] \text{ }.  \label{A19}
\end{equation}%
More complicated structure has function $L$: 
\begin{equation}
L=-8ik_{1}\beta \kappa ^{-1}-2\kappa ^{-1}\bar{k}_{C}\left( \mathbf{G}%
^{\left( 0\right) }\right) ^{C2}-2NQ+R\text{ },  \label{A20}
\end{equation}%
where $R$ is the real function which has to be calculated from the following
quadrature: 
\begin{equation}
R_{,\mu }=4\kappa ^{-1}\bar{k}_{A}\left( \mathbf{\bar{U}}_{\mu }^{\left(
0\right) }\right) _{1}{}^{A}+4\bar{\kappa}^{-1}k_{A}\left( \mathbf{U}_{\mu
}^{\left( 0\right) }\right) _{1}{}^{A}\text{ }.  \label{A21}
\end{equation}%
By the direct calculations it is easy to show that integrability conditions
of equations (\ref{A17}) and (\ref{A19}) for functions $\kappa ^{-1}$ and $Q$
are guaranteed by the fact that matrices $\mathbf{U}_{\mu }$ defined by
formula (\ref{I25}) automatically satisfied the identity $\epsilon ^{\mu \nu
}\mathbf{U}_{\mu ,\nu }=0$ if Einstein-Maxwell equations (\ref{I18}) are
satisfied. This is indeed the case for the background matrices $\mathbf{U}%
_{\mu }^{\left( 0\right) }$ entering the right hand sides of (\ref{A17}) and
(\ref{A19}) since all background quantities represent a solution of the
Einstein-Maxwell equations by definition. Because all other terms in the
right hand sides of (\ref{A17}) and (\ref{A19}) are constants the
integrability conditions of these equations are satisfied evidently.

Not so simple are circumstances about integrability of the equation (\ref%
{A21}) for function $R.$ To prove this integrability we need to show that
quantity $\epsilon ^{\mu \nu }R_{,\mu \nu }$ vanish. Substituting into this
expression the derivative over $x^{\nu }$ of the right hand side of the (\ref%
{A21}) and eliminating from the result derivatives of the functions $\kappa
^{-1}$ and $\bar{\kappa}^{-1}$ using equation (\ref{Dop6}), one can show by
the direct calculation that $\epsilon ^{\mu \nu }R_{,\mu \nu }$ indeed is
zero.

\section{Metric and electromagnetic potentials for the $w$-pole at infinity}

Now from (\ref{A12})-(\ref{A21}) we know the structure of matrix $\mathbf{P}$
and from relation (\ref{A6}) we can find the final solution for the matrix $%
\mathbf{G}$, that is the final expressions for metric coefficient $g_{ab}$
and electromagnetic potentials $\Phi _{a}$ in terms of the given background
solution $g_{ab}^{\left( 0\right) },\Phi _{a}^{\left( 0\right) }$ of the
Einstein-Maxwell equations. The result for the metric is: 
\begin{equation}
g_{11}=\left( \kappa \bar{\kappa}\right) ^{-1}g_{11}^{\left( 0\right)
}+2R\left( k_{1}g_{12}^{\left( 0\right) }-k_{2}g_{11}^{\left( 0\right)
}\right) +R^{2}\kappa \bar{\kappa}\left( k_{2}^{2}g_{11}^{\left( 0\right)
}+k_{1}^{2}g_{22}^{\left( 0\right) }-2k_{1}k_{2}g_{12}^{\left( 0\right)
}\right) \text{ },  \label{B1}
\end{equation}%
\begin{equation}
g_{22}=\kappa \bar{\kappa}k_{1}^{-2}\left( k_{2}^{2}g_{11}^{\left( 0\right)
}+k_{1}^{2}g_{22}^{\left( 0\right) }-2k_{1}k_{2}g_{12}^{\left( 0\right)
}\right) \text{ },  \label{B2}
\end{equation}%
\begin{equation}
g_{12}=k_{1}^{-1}\left( k_{1}g_{12}^{\left( 0\right) }-k_{2}g_{11}^{\left(
0\right) }\right) +k_{1}^{-1}R\kappa \bar{\kappa}\left(
k_{2}^{2}g_{11}^{\left( 0\right) }+k_{1}^{2}g_{22}^{\left( 0\right)
}-2k_{1}k_{2}g_{12}^{\left( 0\right) }\right) \text{ }.  \label{B3}
\end{equation}%
For the electromagnetic potentials the resulting expressions are:%
\begin{equation}
\Phi _{1}=\frac{1}{2}\bar{Q}-\frac{1}{2}\kappa R\left( 2k_{1}\Phi
_{2}^{\left( 0\right) }-2k_{2}\Phi _{1}^{\left( 0\right) }+\bar{k}%
_{3}\right) -\frac{1}{\bar{\kappa}}\Phi _{1}^{\left( 0\right) }\text{ },
\label{B4}
\end{equation}

\begin{equation}
\Phi _{2}=-\frac{\kappa }{2k_{1}}\left( 2k_{1}\Phi _{2}^{\left( 0\right)
}-2k_{2}\Phi _{1}^{\left( 0\right) }+\bar{k}_{3}\right) \text{ }.  \label{B5}
\end{equation}

With these results we have from (\ref{I20}) and (\ref{Dop3}) matrices $%
\mathbf{G}$ and $\mathbf{Y}$ after which we can calculate matrices $\mathbf{K%
}_{\mu }$ from (\ref{I23}) and take quadrature for metric coefficient $f$ \
from equation (\ref{I22}). These calculations are straightforward but
lengthy. The final result, however, is very simple:%
\begin{equation}
f=c_{0}(\kappa \bar{\kappa})^{-1}f^{\left( 0\right) },  \label{B6}
\end{equation}%
where $c_{0}$ is an arbitrary real constant and $f^{\left( 0\right) }$
represents the background solution for the metric coefficient $f$.

\section{Case of the flat background}

Let's consider the simplest case for the background solution, that is the
flat space without electromagnetic field. For the background metric we have:%
\begin{equation}
\left( ds^{\left( 0\right) }\right) ^{2}=f^{\left( 0\right) }\eta _{\mu \nu
}dx^{\mu }dx^{\nu }+g_{11}^{\left( 0\right) }\left( dx^{1}\right)
^{2}+g_{22}^{\left( 0\right) }\left( dx^{2}\right) ^{2}\text{ },  \label{F1}
\end{equation}%
where $\eta _{\mu \nu }$ is the same as in (\ref{I2}) and metric
coefficients are: 
\begin{equation}
\text{ }g_{11}^{\left( 0\right) }=e\text{ },\text{ \ }g_{22}^{\left(
0\right) }=\alpha ^{2},\text{ \ }f^{\left( 0\right) }=-c_{1}eD\text{ },
\label{F2}
\end{equation}%
where $\alpha $ is some solution of the equation (\ref{I8}), $c_{1}$ is an
arbitrary real constant and $D$ has been defined by the formula (\ref{I23}).
The background electromagnetic potentials are absent\footnote{%
The fact that (\ref{F1})-(\ref{Dop7}) represent the solution of
Einstein-Maxwell equations for the flat space-time can be seen easily if we
pass from the coordinates $x^{\mu }$ to the pair of coordinates $\alpha $
and $\beta $. Using definition (\ref{I9}) for function $\beta $ we can
derive the identity: 
\begin{equation*}
-ed\alpha ^{2}+d\beta ^{2}=-eD\eta _{\mu \nu }dx^{\mu }dx^{\nu }.
\end{equation*}%
Due to this relation metric (\ref{F1})-(\ref{F2}) can be written in the
form: 
\begin{equation*}
ds^{2}=c_{1}\left( -ed\alpha ^{2}+d\beta ^{2}\right) +e\left( dx^{1}\right)
^{2}+\alpha ^{2}\left( dx^{2}\right) ^{2}\text{ },
\end{equation*}%
which is evidently flat. Using matrices $\mathbf{G}^{\left( 0\right) }$ and $%
\mathbf{Y}^{\left( 0\right) }$ from (\ref{F3}) it is easy to calculate
matrices $\mathbf{K}_{\mu }^{\left( 0\right) }$ from (\ref{I23}) and
substitute them into the equation (\ref{I22}) to check that this equation
for coefficient $f^{\left( 0\right) }$ is automatically satisfied.}:%
\begin{equation}
\Phi _{a}^{\left( 0\right) }=0.  \label{Dop7}
\end{equation}%
First of all we need the background matrices $\mathbf{G}^{\left( 0\right) },$
$\mathbf{Y}^{\left( 0\right) }$and $\mathbf{U}_{\mu }^{\left( 0\right) }$.
These follow from formulas (\ref{I20}), (\ref{Dop3}), (\ref{I25})
respectively:%
\begin{equation}
\mathbf{G}^{\left( 0\right) }=\left( 
\begin{array}{ccc}
4\alpha ^{2} & 0 & 0 \\ 
0 & 4e & 0 \\ 
0 & 0 & 1%
\end{array}%
\right) \text{ },\text{ }\mathbf{Y}^{\left( 0\right) }=\left( 
\begin{array}{ccc}
1 & 0 & 0 \\ 
0 & 1 & 0 \\ 
0 & 0 & 2%
\end{array}%
\right) \text{ },\text{ }\mathbf{U}_{\mu }^{\left( 0\right) }=\left( 
\begin{array}{ccc}
-2i\beta _{,\mu } & 0 & 0 \\ 
-2\alpha \alpha _{,\mu } & 0 & 0 \\ 
0 & 0 & 0%
\end{array}%
\right) \text{ }.  \label{F3}
\end{equation}%
Although the flat background spectral matrix $\varphi ^{(0)}(x^{\mu },w)$
satisfying equation (\ref{I24}) with $\mathbf{U}_{\mu }^{\left( 0\right) }$
from (\ref{F3}) we need not in this section, it would be reasonable to show
it since we will need it later. This matrix has the following structure:%
\begin{equation}
\mathbf{\varphi }^{(0)}=\left( 
\begin{array}{ccc}
v^{-1} & 0 & 0 \\ 
ie\left( w+\beta \right) v^{-1} & 1 & 0 \\ 
0 & 0 & 1%
\end{array}%
\right) \text{ },\text{ \ }\left[ \mathbf{\varphi }^{(0)}\right]
^{-1}=\left( 
\begin{array}{ccc}
v & 0 & 0 \\ 
-ie\left( w+\beta \right) & 1 & 0 \\ 
0 & 0 & 1%
\end{array}%
\right) \text{ },  \label{F3-A}
\end{equation}%
where function $v$ is defined by the relation:%
\begin{equation}
v^{2}=\left( w+\beta \right) ^{2}-e\alpha ^{2}.  \label{F3-B}
\end{equation}

The second step is to calculate functions $\kappa ,$ $Q,$ $R$ from equations
(\ref{A16}), (\ref{A17}), (\ref{A19}), (\ref{A21}), in the right hand side
of which we should use the background values (\ref{F3}) for the
corresponding matrices. All three quadratures can be taking easily and gives
the following explicit expressions:%
\begin{equation}
\kappa ^{-1}=4k_{1}^{2}\alpha ^{2}+4ek_{2}^{2}+1+i(c_{2}-8k_{1}k_{2}\beta )%
\text{ },  \label{F4}
\end{equation}
\begin{equation}
Q=-8ik_{1}k_{3}\beta +c_{3}+ic_{4}\text{ },  \label{F5}
\end{equation}%
\begin{equation}
R=-16k_{1}c_{2}\beta +64k_{1}^{2}k_{2}\beta ^{2}+c_{5}\text{ },  \label{F6}
\end{equation}%
where $c_{2},$ $c_{3},$ $c_{4},$ $c_{5}$ are new arbitrary real constants.
The potentials $\Phi _{a}$ follow from equations (\ref{B4}) and (\ref{B5})
under condition $\Phi _{a}^{\left( 0\right) }=0$:%
\begin{equation}
\Phi _{1}=\frac{1}{2}\bar{Q}-\frac{1}{2}\bar{k}_{3}\kappa R\text{ },\text{ }%
\Phi _{2}=-\frac{1}{2k_{1}}\bar{k}_{3}\kappa \text{ }.  \label{F7}
\end{equation}%
The real potential of interest $A_{a}=\func{Re}\Phi _{a}$ following from the
last formulas are:%
\begin{equation}
A_{1}=\frac{1}{4}\left( Q+\bar{Q}\right) -\frac{1}{4}R\left( \bar{k}%
_{3}\kappa +k_{3}\bar{\kappa}\right) \text{ },\text{ }A_{2}=-\frac{1}{4k_{1}}%
\left( \bar{k}_{3}\kappa +k_{3}\bar{\kappa}\right) \text{ }.  \label{F8}
\end{equation}%
Substituting into these formulas functions $\kappa $ and $Q$ from (\ref{F4})
and (\ref{F5}) and taking into account that $k_{3}=\cos \gamma +i\sin \gamma
,$ we obtain the following expressions for the electromagnetic potentials:%
\begin{equation}
A_{1}=4k_{1}\beta \sin \gamma +\frac{1}{2}c_{3}+k_{1}RA_{2}\text{ },
\label{F9}
\end{equation}%
\begin{equation}
A_{2}=-\frac{1}{2k_{1}}\frac{\left( 4k_{1}^{2}\alpha
^{2}+4ek_{2}^{2}+1\right) \cos \gamma -\left( c_{2}-8k_{1}k_{2}\beta \right)
\sin \gamma }{\left( 4k_{1}^{2}\alpha ^{2}+4ek_{2}^{2}+1\right) ^{2}+\left(
c_{2}-8k_{1}k_{2}\beta \right) ^{2}}\text{ }.  \label{F10}
\end{equation}

Metric for this solution follows from formulas (\ref{B1})-(\ref{B3}) and (%
\ref{B6}) into which we have to substitute background values $g_{11}^{\left(
0\right) },$ $g_{22}^{\left( 0\right) },$ \ $f^{\left( 0\right) }$ from (\ref%
{F2}). The result is:%
\begin{equation}
g_{11}=e(\kappa \bar{\kappa})^{-1}-2k_{2}eR+\kappa \bar{\kappa}%
(k_{2}^{2}e+k_{1}^{2}\alpha ^{2})R^{2}\text{ },  \label{F11}
\end{equation}%
\begin{equation}
g_{22}=\kappa \bar{\kappa}k_{1}^{-2}\left( k_{2}^{2}e+k_{1}^{2}\alpha
^{2}\right) \text{ },  \label{F12}
\end{equation}%
\begin{equation}
g_{12}=-k_{2}k_{1}^{-1}e+\kappa \bar{\kappa}k_{1}^{-1}\left(
k_{2}^{2}e+k_{1}^{2}\alpha ^{2}\right) R\text{ },  \label{F13}
\end{equation}%
\begin{equation}
f=-c_{0}c_{1}e(\kappa \bar{\kappa})^{-1}D\text{ },  \label{F14}
\end{equation}%
where%
\begin{equation}
\kappa \bar{\kappa}=\left[ \left( 4k_{1}^{2}\alpha ^{2}+4ek_{2}^{2}+1\right)
^{2}+\left( c_{2}-8k_{1}k_{2}\beta \right) ^{2}\right] ^{-1}\text{ },
\label{F15}
\end{equation}%
and function $R$ has been defined in (\ref{F6}). It is easy to check that
from (\ref{F11})-(\ref{F13}) automatically follows relation (\ref{I3}) as it
should be.

Covariant components of electromagnetic tensor follow from (\ref{I5}) for
which we need to calculate the gradients $A_{1,\mu }$ and $A_{2,\mu }$.
These calculations are straightforward and gives the following results:%
\begin{equation}
A_{1,\mu }=\frac{8k_{1}\left( \lambda _{2}\lambda _{1}\cos \gamma -\lambda
_{2}^{2}\sin \gamma \right) \beta _{,\mu }}{\lambda _{1}^{2}+\lambda _{2}^{2}%
}+4k_{1}\beta _{,\mu }\sin \gamma +Rk_{1}A_{2,\mu }\text{ },  \label{F16}
\end{equation}%
\begin{eqnarray}
A_{2,\mu } &=&\frac{4\left[ \left( \lambda _{1}^{2}-\lambda _{2}^{2}\right)
k_{1}\alpha \alpha _{,\mu }-2\lambda _{1}\lambda _{2}k_{2}\beta _{,\mu }%
\right] \cos \gamma }{\left( \lambda _{1}^{2}+\lambda _{2}^{2}\right) ^{2}}-
\label{F17} \\
&&\frac{4\left[ \left( \lambda _{1}^{2}-\lambda _{2}^{2}\right) k_{2}\beta
_{,\mu }+2\lambda _{1}\lambda _{2}k_{1}\alpha \alpha _{,\mu }\right] \sin
\gamma }{\left( \lambda _{1}^{2}+\lambda _{2}^{2}\right) ^{2}}  \notag
\end{eqnarray}%
where%
\begin{equation}
\lambda _{1}=4k_{1}^{2}\alpha ^{2}+4ek_{2}^{2}+1\text{ },\text{ }\lambda
_{2}=c_{2}-8k_{1}k_{2}\beta \text{ }.  \label{F18}
\end{equation}

\section{Melvin magnetic universe as Einstein-Maxwell soliton}

From the previous section it easy to show that the well known Melvin
magnetic universe represents nothing else but solitonic configuration
corresponding to the pole at infinity of the complex plane of the spectral
parameter. To see this it is necessary to take the following particular case
of the general solution derived in the previous section. First of all we
choose the stationary case, that is%
\begin{equation}
e=-1\text{ }.  \label{F19}
\end{equation}%
Then we leave $k_{1}$ to be an arbitrary constant and designate it as $\frac{%
1}{4}B_{0}$ but constants $k_{2}$ and $\gamma $ we put to zero: 
\begin{equation}
k_{1}=\frac{1}{4}B_{0},\text{ }k_{2}=0,\text{ }\gamma =0\text{ (that is }%
k_{3}=1\text{) }.  \label{F20}
\end{equation}%
Also we eliminate arbitrary constant parameters $c_{2}-c_{5}$:%
\begin{equation}
c_{2}=c_{3}=c_{4}=c_{5}=0\text{ }.  \label{F21}
\end{equation}%
The coordinates $x^{1}$ and $x^{2}$ we can understand as time $t$ and
azimuthal angle $\varphi $ of cylindrical coordinate system and for the
stationary case ($e=-1$) we always can choose functions $\alpha $ and $\beta 
$ [which are solutions of equations (\ref{I8})-(\ref{I9})] to be the radial
cylindrical coordinate ($\alpha =\rho $) and coordinate along the axis of
symmetry ($\beta =-z$):%
\begin{equation}
(\alpha ,\beta ,x^{1},x^{2})=(\rho ,-z,t,\varphi )\text{ }  \label{F22}
\end{equation}%
(minus in front of $z$ we choose only for technical convenience, as can be
seen from (\ref{I9}) this choice corresponds to the following definition of
the 2-dimensional greek antisymmetric symbol $\epsilon ^{\mu \nu }$: $%
\epsilon ^{\rho z}=1,$ $\epsilon ^{z\rho }=-1$). With such choice of
arbitrary constants and coordinates designation the functions $\kappa ,Q,R$
following from (\ref{F4})-(\ref{F6}) are:%
\begin{equation}
\kappa ^{-1}=1+\frac{1}{4}B_{0}^{2}\rho ^{2},\text{ }Q=2iB_{0}z,\text{ }R=0%
\text{ }.  \label{F23}
\end{equation}%
As for the arbitrary constants $c_{0}$ and $c_{1}$ in (\ref{F14}) we can
take them to be unities without loss of generality:%
\begin{equation}
c_{0}=c_{1}=1\text{ }.  \label{F24}
\end{equation}%
The factor $D$ in (\ref{F14}) under the choice $\alpha =\rho $ also becomes
unity:%
\begin{equation*}
D=1\text{ }.
\end{equation*}%
Now the metric follows from (\ref{F11})-(\ref{F15}):%
\begin{equation}
ds^{2}=(1+\frac{1}{4}B_{0}^{2}\rho ^{2})^{2}(d\rho ^{2}+dz^{2}-dt^{2})+(1+%
\frac{1}{4}B_{0}^{2}\rho ^{2})^{-2}\rho ^{2}d\varphi ^{2}\text{ },
\label{F25}
\end{equation}%
and electromagnetic potentials from (\ref{F9})-(\ref{F10}):%
\begin{equation}
A_{t}=0,\text{ }A_{\varphi }=-\frac{2}{B_{0}}(1+\frac{1}{4}B_{0}^{2}\rho
^{2})^{-1}\text{ }.  \label{F26}
\end{equation}%
The only non-vanishing component of the electromagnetic tensor is:%
\begin{equation}
F_{\rho \varphi }=-F_{\varphi \rho }=B_{0}\rho (1+\frac{1}{4}B_{0}^{2}\rho
^{2})^{-2}\text{ }.  \label{F27}
\end{equation}%
Complex electromagnetic potential (\ref{I14}) for the Melvin solution
follows from (\ref{F7}) and (\ref{F23}):%
\begin{equation}
\Phi _{t}=-iB_{0}z\text{ },\text{ \ }\Phi _{\varphi }=-\frac{2}{B_{0}}(1+%
\frac{1}{4}B_{0}^{2}\rho ^{2})^{-1}\text{ }.  \label{F27-A}
\end{equation}%
The limit $B_{0}\rightarrow 0$ gives (after adding to the potential $%
A_{\varphi }$ an appropriate calibration constant) flat space-time without
electromagnetic field.

\section{Black hole immersed into Melvin universe}

Using the generating technique outlined in the subsection 1.2 (and in the
subsection 3.3.4 of the book \cite{BV}) it is straightforward to immerse a
generalized black hole (with rotation, electric charge, magnetic charge and
NUT parameter) into the Melvin magnetic universe. Technically this can be
done in three different way. The first one is to add to the flat Minkowski
space-time, choosing it as background, two solitons one of which corresponds
to the pole in the finite region of the $w$-complex plane and other is
located at infinity of this plane. Such approach would corresponds to the
dressing matrix we mentioned in Introduction by the formula (\ref{V3}).
Second way would be to choose as background the well known exact solution
for generalized black hole and introduce on it the solitonic perturbation
with pole at infinity of the $w$-plane. The third possibility is to choose
for the background the Melvin magnetic geon derived in the previous section
and add to this background one soliton corresponding to the finite pole in
the complex plane of spectral parameter. Each way has its technical
advantages and complications. Let's choose the third approach which on our
opinion is more straightforward.

So then we consider the Melvin solution (\ref{F25})-(\ref{F26}) as new
background and all quantities corresponding this solution we designate now
by index "M" (in order to distinguish them from flat background quantities
for which we used before the index "0"). Then we have%
\begin{equation}
\lbrack ds^{\left( M\right) }]^{2}=f^{\left( M\right) }(d\rho
^{2}+dz^{2})+g_{tt}^{\left( M\right) }dt^{2}+g_{\varphi \varphi }^{\left(
M\right) }d\varphi ^{2}\text{ },  \label{F28}
\end{equation}%
\begin{equation}
g_{tt}^{\left( M\right) }=-(1+\frac{1}{4}B_{0}^{2}\rho ^{2})^{2},\text{ }%
g_{\varphi \varphi }^{\left( M\right) }=\rho ^{2}(1+\frac{1}{4}B_{0}^{2}\rho
^{2})^{-2},\text{ }f^{(M)}=(1+\frac{1}{4}B_{0}^{2}\rho ^{2})^{2}\text{ },
\label{F29}
\end{equation}%
\begin{equation}
A_{t}^{(M)}=0,\text{ }A_{\varphi }^{(M)}=-\frac{2}{B_{0}}(1+\frac{1}{4}%
B_{0}^{2}\rho ^{2})^{-1}\text{ },  \label{F30}
\end{equation}%
\begin{equation}
\Phi _{t}^{(M)}=-iB_{0}z\text{ },\text{ \ }\Phi _{\varphi }^{(M)}=-\frac{2}{%
B_{0}}(1+\frac{1}{4}B_{0}^{2}\rho ^{2})^{-1}\text{ },  \label{F31}
\end{equation}%
and we should keep in mind the choosed numerics for the mute Latin
coordinates $x^{a}$, that is $(x^{1},x^{2})=(t,\varphi )$.

First of all we need the Melvin background spectral matrix $\mathbf{\varphi }%
^{(M)}(\rho ,z,w)$ which is appropriately normalized solution of equation (%
\ref{I24}) applied to the Melvin metric and electromagnetic potentials (\ref%
{F28})-(\ref{F31}). In general to integrate equation (\ref{I24}) is not a
simple task. Fortunately in our present problem this matrix is already
known. As follows from (\ref{I27}) and (\ref{A1})-(\ref{A2}) it is%
\begin{equation}
\mathbf{\varphi }^{(M)}=(\mathbf{P}^{(M)}+\mathbf{A}^{(M)}w)\mathbf{\varphi }%
^{(0)}\text{ },  \label{F32}
\end{equation}%
\begin{equation}
\left[ \mathbf{\varphi }^{(M)}\right] ^{-1}=\left[ \mathbf{\varphi }^{(0)}%
\right] ^{-1}\left\{ \left[ \mathbf{P}^{(M)}\right] ^{-1}-\left[ \mathbf{P}%
^{(M)}\right] ^{-1}\mathbf{A}^{(M)}\left[ \mathbf{P}^{(M)}\right]
^{-1}w\right\} \text{ },  \label{F33}
\end{equation}%
where $\mathbf{\varphi }^{(0)}$ is the flat background spectral matrix (\ref%
{F3-A}) which for the stationary case and choosed functions $\alpha ,\beta $
[see (\ref{F22})] is: 
\begin{equation}
\mathbf{\varphi }^{(0)}=\left( 
\begin{array}{ccc}
v^{-1} & 0 & 0 \\ 
-i\left( w-z\right) v^{-1} & 1 & 0 \\ 
0 & 0 & 1%
\end{array}%
\right) \text{ },\text{ \ }\left[ \mathbf{\varphi }^{(0)}\right]
^{-1}=\left( 
\begin{array}{ccc}
v & 0 & 0 \\ 
i\left( w-z\right) & 1 & 0 \\ 
0 & 0 & 1%
\end{array}%
\right) \text{ },  \label{F34}
\end{equation}%
where%
\begin{equation}
v^{2}=\left( w-z\right) ^{2}+\rho ^{2}.  \label{F35}
\end{equation}%
Matrices $\mathbf{P}^{(M)},\left[ \mathbf{P}^{(M)}\right] ^{-1}$and $\mathbf{%
A}^{(M)}$ follow from formulas (\ref{A11})-(\ref{A13}) for those particular
values of all constant parameters and functions which correspond to the
Melvin solution and which was specified in the section "Melvin magnetic
universe as Einstein-Maxwell soliton". In addition to the all quantities of
this section we need also functions $N$ and $L$ defined by (\ref{A18}) and (%
\ref{A20}) which functions also enter the matrices $\mathbf{P}^{(M)}$and $%
\left[ \mathbf{P}^{(M)}\right] ^{-1}$. It is easy to see that for the Melvin
background these two quantities are:%
\begin{equation}
N=1,\text{ }L=-2iB_{0}\left( 1-\frac{1}{4}B_{0}^{2}\rho ^{2}\right) .
\label{F36}
\end{equation}%
Now we have everything to write down matrices $\mathbf{P}^{(M)},\left[ 
\mathbf{P}^{(M)}\right] ^{-1},\mathbf{A}^{(M)}$. The result is:%
\begin{equation}
\mathbf{A}^{(M)}\mathbf{=}\left( 
\begin{array}{ccc}
0 & -\frac{1}{2}iB_{0}^{2} & 0 \\ 
0 & 0 & 0 \\ 
0 & 0 & 0%
\end{array}%
\right) \text{ },  \label{F37}
\end{equation}%
\begin{equation}
\mathbf{P}^{(M)}\mathbf{=}\left( 
\begin{array}{ccc}
1-\frac{1}{4}B_{0}^{2}\rho ^{2} & \frac{1}{2}iB_{0}^{2}z & -\frac{1}{2}B_{0}
\\ 
0 & \left( 1+\frac{1}{4}B_{0}^{2}\rho ^{2}\right) ^{-1} & 0 \\ 
\frac{4}{B_{0}} & 2iB_{0}z\left( 1+\frac{1}{4}B_{0}^{2}\rho ^{2}\right) ^{-1}
& -1%
\end{array}%
\right) \text{ },  \label{F38}
\end{equation}%
\begin{equation}
\lbrack \mathbf{P}^{(M)}]^{-1}\mathbf{=}\left( 1+\frac{1}{4}B_{0}^{2}\rho
^{2}\right) ^{-1}\left( 
\begin{array}{ccc}
-1 & -\frac{1}{2}iB_{0}^{2}z\left( 1-\frac{1}{4}B_{0}^{2}\rho ^{2}\right) & 
\frac{1}{2}B_{0} \\ 
0 & \left( 1+\frac{1}{4}B_{0}^{2}\rho ^{2}\right) ^{2} & 0 \\ 
-\frac{4}{B_{0}} & iB_{0}^{3}z\rho ^{2} & 1-\frac{1}{4}B_{0}^{2}\rho ^{2}%
\end{array}%
\right) \text{ }.  \label{F39}
\end{equation}%
Substituting these expressions into (\ref{F32})-(\ref{F33}) we obtain the
final result for the Melvin spectral matrix:%
\begin{equation}
\mathbf{\varphi }^{(M)}=\left( 
\begin{array}{ccc}
1-\frac{1}{4}B_{0}^{2}\rho ^{2} & -\frac{1}{2}iB_{0}^{2}(w-z) & -\frac{1}{2}%
B_{0} \\ 
0 & \left( 1+\frac{1}{4}B_{0}^{2}\rho ^{2}\right) ^{-1} & 0 \\ 
\frac{4}{B_{0}} & 2iB_{0}z\left( 1+\frac{1}{4}B_{0}^{2}\rho ^{2}\right) ^{-1}
& -1%
\end{array}%
\right) \mathbf{\varphi }^{(0)}\text{ },  \label{F40}
\end{equation}%
\begin{gather}
\left[ \mathbf{\varphi }^{(M)}\right] ^{-1}=\left( 1+\frac{1}{4}%
B_{0}^{2}\rho ^{2}\right) ^{-1}\left[ \mathbf{\varphi }^{(0)}\right]
^{-1}\times  \label{F41} \\
\left( 
\begin{array}{ccc}
-1 & -\frac{1}{2}iB_{0}^{2}\left[ w+z+\frac{1}{4}B_{0}^{2}\rho ^{2}\left(
w-z\right) \right] & \frac{1}{2}B_{0} \\ 
0 & \left( 1+\frac{1}{4}B_{0}^{2}\rho ^{2}\right) ^{2} & 0 \\ 
-\frac{4}{B_{0}} & iB_{0}^{3}z\rho ^{2}-2iB_{0}\left( 1+\frac{1}{4}%
B_{0}^{2}\rho ^{2}\right) w & 1-\frac{1}{4}B_{0}^{2}\rho ^{2}%
\end{array}%
\right)  \notag
\end{gather}

Formulas (\ref{F40})-(\ref{F41}) for the background spectral matrix are most
fundamental because all the rest is just algebraic manipulations. Following
the instructions given in subsection 3.3.4 of the book \cite{BV} we should
calculate three-dimensional vectors $p_{A}$ and $m^{A}$. If the pole in $w$%
-plane is located in some finite point $w=w_{1}$ these vectors are:%
\begin{equation}
p_{A}=l^{B}\left[ \mathbf{\varphi }^{(M)}(w_{1})\right] _{AB}\text{ },
\label{F42}
\end{equation}%
\begin{equation}
m^{A}=C_{BD}\bar{l}^{D}\left[ \left( \mathbf{\varphi }^{(M)}\right)
^{-1}\left( w_{1}\right) \right] ^{BA}\text{ },  \label{F43}
\end{equation}%
\begin{equation}
C_{BD}=diag\left( 4,-4,1\right) \text{ },  \label{F44}
\end{equation}%
where $\mathbf{\varphi }^{(M)}(w_{1})$ and $\left( \mathbf{\varphi }%
^{(M)}\right) ^{-1}\left( w_{1}\right) $ are the matrices (\ref{F40})-(\ref%
{F41}) calculated at the point $w=w_{1}$ and $l^{A}$ are three arbitrary
complex constants. In constants $l^{A}$ and $w_{1}$ are encoded all
arbitrary parameters of the black hole we immersed in the Melvin universe.
In fact the final metric and electromagnetic potentials does not depend
separately on the constant $l^{1}$, they depend only on the ratios $%
l^{2}/l^{1}$ and $l^{3}/l^{1}$. These two arbitrary complex ratios together
with pole parameter $w_{1}$ generate six real parameters of black hole: mass
and NUT parameter (ratio $l^{2}/l^{1}$), electric and magnetic charge (ratio 
$l^{3}/l^{1}$), angular momentum ($\func{Im}w_{1}$) and position of the
black hole on the symmetry axis ($\func{Re}w_{1}$).

Final solution for the metric and electromagnetic potentials, that is for
the matrix $\mathbf{G}$ (\ref{I20}), can be expressed in terms of the matrix 
$\mathbf{S}$ (with components $\left( \mathbf{S}\right) _{A}{}^{B}$) (see
subsection 3.3.3 in the book \cite{BV}):%
\begin{equation}
\left( \mathbf{S}\right) _{A}{}^{B}=-\left( w_{1}-\bar{w}_{1}\right) \left(
p_{C}m^{C}\right) ^{-1}p_{A}m^{B}\text{ },  \label{F45}
\end{equation}%
by the formula 
\begin{equation}
\mathbf{G}=\mathbf{G}^{(M)}+4i\left( \mathbf{S}^{\dagger }\mathbf{\Omega
+\Omega S}\right) \text{ }.  \label{F46}
\end{equation}%
where $\mathbf{G}^{(M)}$ is background matrix $\mathbf{G}$ (\ref{I20}) for
the Melvin solution and matrix $\mathbf{\Omega }$\textbf{\ }was defined in (%
\ref{I26}). Due to the simple structure of matrix $\mathbf{\Omega }$ the
formula (\ref{F46}) can be written in the form:%
\begin{equation}
\mathbf{G}=\mathbf{G}^{(M)}+4i\left( 
\begin{array}{ccc}
\left( \mathbf{S}\right) _{2}{}^{1}-\left( \mathbf{\bar{S}}\right) _{2}{}^{1}
& \left( \mathbf{S}\right) _{2}{}^{2}+\left( \mathbf{\bar{S}}\right)
_{1}{}^{1} & \left( \mathbf{S}\right) _{2}{}^{3} \\ 
-\left( \mathbf{S}\right) _{1}{}^{1}-\left( \mathbf{\bar{S}}\right)
_{2}{}^{2} & -\left( \mathbf{S}\right) _{1}{}^{2}+\left( \mathbf{\bar{S}}%
\right) _{1}{}^{2} & -\left( \mathbf{S}\right) _{1}{}^{3} \\ 
-\left( \mathbf{\bar{S}}\right) _{2}{}^{3} & \left( \mathbf{\bar{S}}\right)
_{1}{}^{3} & 0%
\end{array}%
\right) \text{ }.  \label{F47}
\end{equation}%
The conformal factor $f$ for the final solution follows from the section 3.4
of the book \cite{BV} and it is:%
\begin{equation}
f=c_{0}\frac{p_{A}m^{A}\bar{p}_{B}\bar{m}^{B}}{\left( \func{Im}w_{1}\right)
^{2}}f^{\left( M\right) }  \label{F48}
\end{equation}%
where $c_{0}$ is an arbitrary real constant and $f^{\left( M\right) }$ is
the Melvin background value of this factor defined in the formula (\ref{F29}%
).

There is no big sense to express this final solution in closed analytical
form in terms of elementary functions on coordinates $\rho ,z,$ (although
this is possible, of course) because such form is too cumbersome. It is more
convenient to investigate each particular case of interest starting from the
corresponding particular form of Melvin spectral matrix (\ref{F40})-(\ref%
{F41}) and derive all the rest from (\ref{F42})-(\ref{F48}) keeping
calculations strictly in framework of this particular case.

There is further more practical development and applications of the theory
presented here but it is still under progress and we hope to expose it later.

\section{Acknowledgments}

I am grateful to the Albert Einstein Institute for Gravitational Physics
(Potsdam) and to the International Solvay Institute for Physics and
Chemistry (Brussels) for collaboration and support where the part of this
research have been done. My personal gratitudes are to Professors H. Nicolai
and M. Henneaux for invitation and warm welcome.

\end{document}